\documentclass[aps,prl,twocolumn,superscriptaddress]{revtex4}
\usepackage{graphics,epsfig}

\begin{document}

\title{Description of the fluctuating colloid-polymer interface}

\author{Edgar M. Blokhuis}

\author{Joris Kuipers}

\affiliation{Colloid and Interface Science, Leiden Institute of Chemistry, Gorlaeus Laboratories,
P.O. Box 9502, 2300 RA Leiden, The Netherlands.}

\author{Richard Vink}

\affiliation{Institute of Theoretical Physics, Georg-August-Universit\"{a}t,
Friedrich-Hund-Platz 1, D-37077 G\"{o}ttingen, Germany.}

\begin{abstract}
To describe the full spectrum of surface fluctuations of the
interface between phase-separated colloid-polymer mixtures
from low scattering vector $q$ (classical capillary wave
theory) to high $q$ (bulk-like fluctuations), one must 
take account of the interface's bending rigidity. We find
that the bending rigidity is negative and that on approach
to the critical point it vanishes proportionally to the
interfacial tension. Both features are in agreement
with Monte Carlo simulations.
\end{abstract}

\maketitle

One of the outstanding theoretical problems in the understanding of
the structure of a simple liquid surface is the description of the full
spectrum of surface fluctuations obtained in light scattering
experiments \cite{Daillant1, Daillant2} and computer simulations
\cite{Stecki, Vink, Tarazona}. Insight into the structure of a simple
liquid surface is provided by molecular theories \cite{Evans, RW},
such as the van der Waals squared-gradient model, on the one hand
and the capillary wave model \cite{BLS, Weeks} on the other hand.
The theoretical challenge is to incorporate both theories and to
describe the spectrum of fluctuations of a liquid surface from the
molecular scale to the scale of capillary waves.

Here, we report on a theoretical description of Monte Carlo (MC)
simulations \cite{Vink} of a system consisting of a mixture of
colloidal particles with diameter $d$ and polymers with a radius
of gyration $R_{\rm g}$.
The presence of polymer induces a depletion attraction \cite{AOVrij}
between the colloidal particles which may ultimately induce phase
separation \cite{Gast, Lekkerkerker}. The resulting interface of the
demixed colloid-polymer system is studied for a number of polymer
concentrations and for a polymer-colloid size ratio
$\varepsilon \!\equiv\! 1 + 2 R_{\rm g}/d\!=\!$ 1.8.

\begin{figure}
\centering
\includegraphics[angle=270,width=200pt]{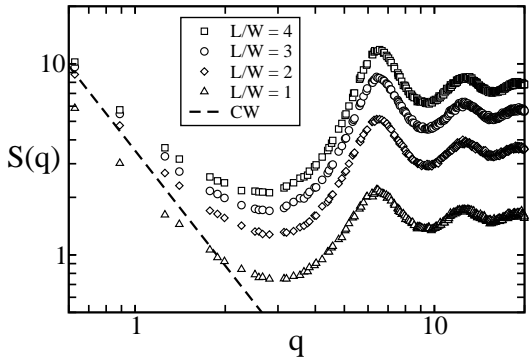}
\caption{MC simulation results for the surface structure factor
(in units of $d^4$) versus $q$ (in units of $1/d$) for various values
of the integration limit $L/W\!=\!$ 1, 2, 3, 4 \cite{Vink}. The dashed
line is the capillary wave model. In this example $\varepsilon\!=\!$ 1.8,
$\eta_{p}\!=\!$ 1.0.}
\label{fig:Vink0}
\end{figure}

The quantity studied in the simulations is the (surface)
density-density correlation function:
\begin{eqnarray}
\label{eq:S(r)}
&& S(r_{\parallel}) \equiv \frac{1}{(\rho_{\ell} - \rho_v)^2} \int\limits_{-L}^{L} \!\! dz_1
\int\limits_{-L}^{L} \!\! dz_2 \\
&& \hspace*{5pt} <\! [ \rho(\vec{r}_1) - \rho_{\rm step}(z_1) ] \,
[ \rho(\vec{r}_2) - \rho_{\rm step}(z_2) ] \!> \,, \nonumber
\end{eqnarray}
where $\rho(\vec{r})$ is the colloidal density, $\vec{r}_{\parallel} \!=\! (x,y)$
is the direction parallel to the surface, and where we have defined
$\rho_{\rm step}(z)\!\equiv\!\rho_{\ell} \, \Theta(-z) + \rho_v \, \Theta(z)$
with $\Theta(z)$ the Heaviside function and $\rho_{\ell,v}$ the bulk
density in the liquid and vapor region, respectively, where by
``liquid'' we mean the phase relatively rich in colloids and by
``vapor'' the phase relatively poor in colloids.
Its Fourier transform is termed the {\em surface structure factor}
\begin{equation}
\label{eq:S(q)}
S(q) = \int \!\! d\vec{r}_{\parallel} \,
e^{- i \vec{q} \cdot \vec{r}_{\parallel}} \, S(r_{\parallel}) \,.
\end{equation}
In Figure \ref{fig:Vink0}, MC simulation results \cite{Vink}
for $S(q)$ are shown for various values of the integration
limit $L$. The Figure shows that the contribution to $S(q)$
from short wavelength fluctuations (high $q$) increases with $L$.

To analyze $S(q)$, one needs to model the density fluctuations in the interfacial
region. In the capillary wave model (CW) \cite{BLS}, the fluctuating
interface is described in terms of a two-dimensional surface height function
$h(\vec{r}_{\parallel})$
\begin{equation}
\label{eq:rho_CW}
\rho(\vec{r}) = \rho_0(z) - \rho_0^{\prime}(z) \, h(\vec{r}_{\parallel}) + \ldots
\end{equation}
where $\rho_0(z)\!= \; <\!\rho(\vec{r})\!>$.
In the {\em extended} capillary wave model (ECW), the expansion in
gradients of $h(\vec{r}_{\parallel})$ is continued \cite{Parry, Blokhuis99, Mecke}:
\begin{equation}
\label{eq:rho_ECW}
\rho(\vec{r}) = \rho_0(z) - \rho^{\prime}_0(z) \, h(\vec{r}_{\parallel})
- \frac{\rho_1(z)}{2} \, \Delta h(\vec{r}_{\parallel}) + \ldots
\end{equation}
The function $\rho_1(z)$ is identified as the correction to the
density profile due to the {\em curvature} of the interface,
$\Delta h(\vec{r}_{\parallel} )\!\approx\! -1/R_1 - 1/R_2$,
with $R_1$ and $R_2$ the (principal) radii of curvature. 

With Eq.(\ref{eq:rho_ECW}) inserted into Eq.(\ref{eq:S(r)}), we find
that $S(q)$ equals the {\em height-height correlation function},
$S(q) \!=\! S_{hh}(q)$, where
\begin{equation}
\label{eq:S(q)_hh}
S_{hh}(q) \equiv \int \!\! d\vec{r}_{\parallel} \,
e^{- i \vec{q} \cdot \vec{r}_{\parallel}} \,
<\! h(\vec{r}_{1,\parallel}) \, h(\vec{r}_{2,\parallel}) \!> \,.
\end{equation}
Here we have assumed that the location of the interface, as described by
the height function $h(\vec{r}_{\parallel})$, is given by the {\em Gibbs
equimolar surface} \cite{Gibbs}, which gives for $\rho_0(z)$ and $\rho_1(z)$:
\begin{equation}
\label{eq:Gibbs}
\!\!\! \int \!\! dz \left[ \, \rho_0(z) - \rho_{step}(z) \, \right] = 0 \,,
\int \!\! dz \; \rho_1(z) = 0 \,.
\end{equation}
Naturally, other choices are possible \cite{Tarazona} and equally legitimate
as long as they lead to a location of the dividing surface that is
`sensibly coincident' \cite{Gibbs} with the interfacial region.

The height-height correlation function $S_{hh}(q)$ is determined by
considering the free energy $\Delta \Omega$ associated with a surface
fluctuation \cite{BLS, Weeks}. The inclusion of a curvature correction to
the free energy is described by the Helfrich free energy \cite{Helfrich}.
It gives for $\Delta \Omega$
\begin{equation}
\label{eq:Omega_ECW}
\Delta \Omega = \frac{1}{2} \int \!\! \frac{d\vec{q}}{(2 \pi)^2} \;
\sigma(q) \, q^2 \, h(\vec{q}) \, h(-\vec{q}) \,,
\end{equation}
with
\begin{equation}
\label{eq:sigma(q)}
\sigma(q) = \sigma + k \, q^2 + \ldots
\end{equation}
The coefficient $k$ is identified as Helfrich's {\em bending rigidity}
\cite{Helfrich, Meunier}. It is important to realize that the bending
rigidity, defined by Eqs.(\ref{eq:Omega_ECW}) and
(\ref{eq:sigma(q)}), depends on the choice made for the location
of the dividing surface (here: the Gibbs equimolar surface for
the colloid component).

Using Eq.(\ref{eq:Omega_ECW}), the height-height correlation function
can be calculated \cite{Meunier}
\begin{equation}
\label{eq:S(q)_hh_ECW}
S_{hh}(q) = \frac{k_{\rm B} T}{\sigma(q) \, q^2}
= \frac{k_{\rm B} T}{\sigma \, q^2 + k \, q^4 + \ldots} \,.
\end{equation}
Without bending rigidity ($k \!=\! 0$) this is the classical capillary wave
result in the absence of gravity (dashed line in Figure \ref{fig:Vink0}).
When $L$ is sufficiently large, the capillary wave model accurately
describes the behavior of $S(q)$ at low $q$.

To model $S(q)$ in the {\em whole} $q$-range, we also include bulk-like
fluctuations to the density:
\begin{equation}
\label{eq:rho_ECWB}
\rho(\vec{r}) \!=\! \rho_0(z) - \rho^{\prime}_0(z) \, h(\vec{r}_{\parallel})
- \frac{\rho_1(z)}{2} \, \Delta h(\vec{r}_{\parallel}) + \delta \rho_b(\vec{r}) .
\end{equation}
Inserting Eq.(\ref{eq:rho_ECWB}) into Eq.(\ref{eq:S(r)}),
one now finds that
\begin{equation}
\label{eq:S(q)_B}
S(q) = S_{hh}(q) + {\cal N}_L \, S_b(q) \,.
\end{equation}
The second term is derived from an integration into the bulk regions
(to a distance $L$) of the bulk structure factor $S_b(q)$
\begin{equation}
S_b(q) = 1 + \rho_b \int \!\! d\vec{r}_{12} \,
e^{- i \vec{q} \cdot \vec{r}_{12}} \; \left[ \, g(r) - 1 \, \right] \,.
\end{equation}
The density correlation function $g(r)$ differs in either phase,
but here we take for it $g_{\ell}(r)$ of the bulk {\em liquid}.
This approximation may be justified by arguing that close to the
critical point there is no distinction between the two bulk correlation
functions, whereas far from the critical point the contribution
from the bulk vapor can be neglected since $\rho_v\!\approx\!0$.
The error is further reduced by fitting the $L$-dependent
prefactor ${\cal N}_L$ to the limiting behavior of $S(q)$ at
$qd \rightarrow\! \infty$.

\begin{figure}
\centering
\includegraphics[angle=270,width=200pt]{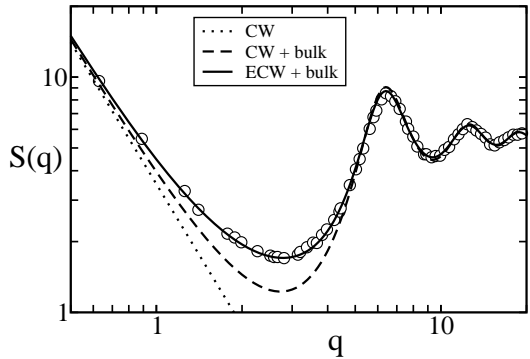}
\caption{MC simulation results \cite{Vink} (circles) for the surface
structure factor (in units of $d^4$) versus $q$ (in units of $1/d$).
The dotted line is the capillary wave model; the dashed line is the
combination of the capillary wave model and the bulk correlation function;
the solid line is the combination of the extended capillary wave model and
the bulk correlation function. In this example $\varepsilon\!=\!$ 1.8,
$\eta_{p}\!=\!$ 1.0, $L/W\!=\!$ 3.}
\label{fig:Vink1}
\end{figure}
In Figure \ref{fig:Vink1}, we show the result from Figure \ref{fig:Vink0}
for $L/W \!=\!$ 3. For $qd \!\ll\! 1$ the results asymptotically
approach the capillary wave model (dotted line). The dashed line is the
result of adding the bulk-like fluctuations to the capillary waves:
\begin{equation}
\label{eq:S(q)_CWB}
S(q) = \frac{k_{\rm B} T}{\sigma q^2} + {\cal N}_L \, S_b(q) \,. 
\end{equation}
Figure \ref{fig:Vink1} shows that Eq.(\ref{eq:S(q)_CWB}) already matches the
simulation results quite accurately except at intermediate values of $q$,
$qd \!\approx\! 1$. 

Finally, we include a bending rigidity in $S(q)$:
\begin{equation}
\label{eq:S(q)_ECWB}
S(q) = \frac{k_{\rm B} T}{\sigma \, q^2 + k \, q^4 + \ldots} + {\cal N}_L \, S_b(q) \,. 
\end{equation}
The value of the bending rigidity is extracted from the behavior
of $S(q)$ at low $q$. The fact that the simulation results in Figure
\ref{fig:Vink1} are systematically {\em above} the capillary wave model in
this region indicates that the bending rigidity thus obtained is {\em negative},
$k \!<\! 0$. Unfortunately, a negative bending rigidity prohibits the
use of $S(q)$ in Eq.(\ref{eq:S(q)_ECWB}) to fit the simulation results in
the {\em entire} $q$-range, since the denominator becomes zero at a certain
value of $q$. It is therefore convenient to rewrite the expansion in $q^2$
in Eq.(\ref{eq:S(q)_ECWB}) as:
\begin{equation}
\label{eq:S(q)_ECWB_new}
S(q) = \frac{k_{\rm B} T}{\sigma q^2} \,
(1 - \frac{k}{\sigma} \, q^2 + \ldots) + {\cal N}_L \, S_b(q) \,, 
\end{equation}
which is equivalent to Eq.(\ref{eq:S(q)_ECWB}) to the order in $q^2$ considered,
but which has the advantage of being well-behaved in the entire $q$-range.
The above form for $S(q)$, with the bending rigidity used as an adjustable
parameter, is plotted in Figure \ref{fig:Vink1} as the solid line.
Exceptionally good agreement with the MC simulations is now obtained for
all $q$. In Table 1, we list the fitted values for the bending rigidity for
a number of different polymer concentrations.

\begin{table}
\centering
\begin{tabular}{c|c|c|c|c|c}
$\eta_{p}$ & $\eta_{\ell}$ & $\eta_v$ & $\sigma$ & $k$ & $\sqrt{-k/\sigma}$ \\
\hline
0.9 & 0.2970 & 0.0141 & 0.1532 & -0.045 (15) & 0.54 \\
1.0 & 0.3271 & 0.0062 & 0.2848 & -0.07 (2)   & 0.50 \\
1.1 & 0.3485 & 0.0030 & 0.4194 & -0.10 (3)   & 0.49 \\
1.2 & 0.3647 & 0.0018 & 0.5555 & -0.14 (3)   & 0.50 \\
\end{tabular}
\caption{MC simulation results \cite{Vink} for the polymer volume fraction $\eta_{p}$,
liquid and vapor colloidal volume fractions, $\eta_{\ell}$ and $\eta_v$,
surface tension $\sigma$ (in units of $k_{\rm B} T / d^2$), bending rigidity $k$
(in units of $k_{\rm B} T$; in parenthesis the estimated error in the last digit),
and $\sqrt{-k/\sigma}$ (in units of $d$).}
\end{table}

\begin{figure}
\centering
\includegraphics[angle=270,width=200pt]{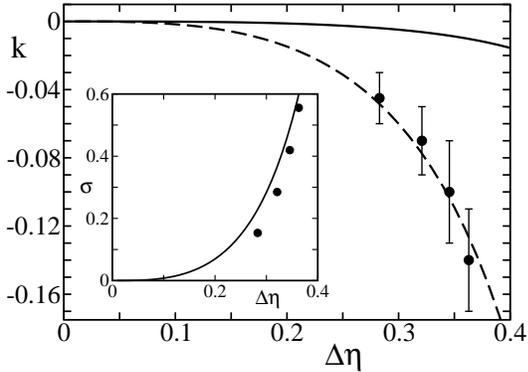}
\caption{Bending rigidity in units of $k_{\rm B} T$
versus the volume fraction difference $\Delta \eta$.
The inset shows the surface tension in units of
$k_{\rm B} T / d^2$. The solid lines are the gradient
expansion approximation; filled circles are the results from
the MC simulations; the dashed line is the fit $\sqrt{-k/\sigma} \approx$~0.47~$d$.}
\label{fig:k_and_sigma}
\end{figure}

Next, we investigate whether the value and behavior of $k$ can be
understood from a molecular theory. One should then consider a
microscopic model for the free energy $\Omega$ to determine the
density profiles $\rho_0(z)$ and $\rho_1(z)$. Here, we consider
the free energy density functional based on a {\em squared-gradient}
expansion \cite{RW, Parry, Blokhuis99, Varea}:
\begin{equation}
\label{eq:Omega}
\Omega[\rho] = \int \!\! d\vec{r} \left[ m \, | \vec{\nabla} \rho(\vec{r}) |^2
- \frac{B}{4} (\Delta \rho(\vec{r}))^2 + g(\rho) \, \right] \,,
\end{equation} 
where the coefficients $m$ and $B$ are defined as
\begin{eqnarray}
\label{eq:m_and_B}
m &\equiv& - \frac{1}{12} \, \int \!\! d\vec{r}_{12} \; r^2 \, U(r) \,, \nonumber \\
B &\equiv& - \frac{1}{60} \, \int \!\! d\vec{r}_{12} \; r^4 \, U(r) \,.
\end{eqnarray}
The integration over $\vec{r}_{12}$ is restricted to the attractive part
($r\!>\!d$) of the interaction potential $U(r)$, for which we consider the
Asakura-Oosawa-Vrij depletion interaction potential \cite{AOVrij}:
\begin{equation}
U(r) = \frac{- k_{\rm B} T \, \eta_p}{2 \, (\varepsilon-1)^3}
\left[ \, 2 \, \varepsilon^3 - 3 \, \varepsilon^2 \left( \frac{r}{d} \right)
+ \left( \frac{r}{d} \right)^{\!3} \right] \,,
\end{equation}
where the intermolecular distance is in the range $1\!<\!r/d\!<\!\varepsilon$.
For explicit calculations, $g(\rho)$ is taken to be of the Carnahan-Starling form:
\begin{equation}
\label{eq:g(rho)}
g(\rho) = k_{\rm B} T \rho \ln(\rho) 
+ k_{\rm B} T \rho \, \frac{(4 \eta - 3 \eta^2)}{(1 - \eta)^2}
- \mu \rho - a \rho^2 \,,
\end{equation} 
where $\eta \!\equiv\! (\pi/6) \, \rho \, d^3$, $\mu \!=\! \mu_{\rm coex}$,
and the van der Waals parameter $a$ is given by
\begin{equation}
\label{eq:a}
a \equiv - \frac{1}{2} \, \int \!\! d\vec{r}_{12} \; U(r) \,.
\end{equation} 
The surface tension, to leading order in the squared-gradient expansion, can be
determined from the usual expression \cite{RW}
\begin{equation}
\label{eq:sigma_SQ}
\sigma = 2 \, \sqrt{m}  \int\limits_{\rho_v}^{\rho_{\ell}} \!\! d\rho \; \sqrt{g(\rho) + p} \,.
\end{equation}
In the inset of Figure \ref{fig:k_and_sigma}, the surface tension is shown
as a function of the colloidal volume fraction difference,
$\Delta \eta \!\equiv\! \eta_{\ell} - \eta_v$. The squared-gradient expression
(solid line) is in satisfactory agreement \cite{Kuipers} with the MC simulations.

The (planar) density profile $\rho_0(z)$ is determined from minimizing
the free energy functional $\Omega[\rho]$ in Eq.(\ref{eq:Omega}) in
planar symmetry. To also determine the density profile $\rho_1(z)$
from a minimization procedure, one should consider the energetically
most favorable density profile for a {\em given} curvature of the surface.
To set the curvature to a specific value, one adds to the free energy in
Eq.(\ref{eq:Omega}) an external field $V_{ext}(\vec{r})$ that acts a
Lagrange multiplier. Different choices for $V_{ext}(\vec{r})$ can then
be made, but we choose it such that it acts only in the interfacial region:
\begin{equation}
\label{eq:V_ext}
V_{ext}(\vec{r}) = \lambda \, \rho^{\prime}_0(z) \, \Delta h(\vec{r}_{\parallel}) \,,
\end{equation}
with the Lagrange multiplier $\lambda$ set by the imposed curvature. This choice
for $V_{ext}(\vec{r})$ constitutes our fundamental
`Ansatz' for the determination of $\rho_1(z)$. It improves on earlier choices made
\cite{Parry, Blokhuis99, Blokhuis93} in the sense that the bulk densities are equal
to those at coexistence and the density profile remains a continuous function.

The minimization of the free energy, with the above external field added,
using the fluctuating density in Eq.(\ref{eq:rho_ECW}) yields the
following Euler-Lagrange (EL) equations for $\rho_0(z)$ and $\rho_1(z)$:
\begin{eqnarray}
\label{eq:EL}
g^{\prime}(\rho_0) &=& 2 m \, \rho^{\prime\prime}_0(z) \,, \nonumber \\
g^{\prime\prime}(\rho_0) \rho_1(z) &=& 2 m \, \rho^{\prime\prime}_1(z)
+ 4 m \, \rho^{\prime}_0(z) \\
&& + 2 B \, \rho^{\prime\prime\prime}_0(z) + 2 \lambda \, \rho^{\prime}_0(z) \,. \nonumber 
\end{eqnarray}
The change in free energy $\Delta \Omega$ due a certain density fluctuation
is determined by inserting $\rho(\vec{r})$ in Eq.(\ref{eq:rho_ECW})
into the expression for $\Omega$ in Eq.(\ref{eq:Omega}). One finds that
$\Delta \Omega$ is then given by the expression in Eq.(\ref{eq:Omega_ECW}),
with the bending rigidity \cite{Blokhuis99}
\begin{equation}
\label{eq:k}
\!\! k = - 2 m \int \!\! dz \; \rho_1(z) \, \rho_0^{\prime}(z)
- \frac{B}{2} \int \!\! dz \; \rho_0^{\prime}(z)^2 \,,
\end{equation}
where we have used the EL equations in Eq.(\ref{eq:EL}).

To determine $\rho_0(z)$ we assume proximity to the {\em critical point}
where $g(\rho)$ takes on the usual double-well form. The solution of
the Euler-Lagrange equation in Eq.(\ref{eq:EL}) then gives \cite{RW}:
\begin{equation}
\label{eq:tanh}
\rho_0(z) = \frac{1}{2} (\rho_{\ell} + \rho_v) - \frac{\Delta \rho}{2} \, \tanh(z/2\xi) \,,
\end{equation}
where $\xi$ is a measure of the interfacial thickness which we shall
define as $\xi \!\equiv\! m \, (\Delta \rho)^2/(3 \, \sigma)$,
with the value of $\sigma$ given by Eq.(\ref{eq:sigma_SQ}).
To determine $\rho_1(z)$ the differential equation in
Eq.(\ref{eq:EL}) is solved using the $\tanh$-profile for $\rho_0(z)$,
yielding:
\begin{equation}
\label{eq:rho1}
\!\!\! \rho_1(z) = \frac{3 \, B}{10 \, m} \frac{\Delta \rho}{\xi} \,
\frac{\left[ \, 1 - \ln(2 \, \cosh(z/2\xi)) \, \right]}{\cosh^2(z/2\xi)} \,,
\end{equation}
where we have used that $\lambda \!=\! - 2 m + B/(5 \xi^2)$.

Inserting Eq.(\ref{eq:rho1}) into Eq.(\ref{eq:k}),
one finds for $k$
\begin{equation}
\label{eq:k_res}
k = - \frac{B \, (\Delta \rho)^2}{60 \, \xi} = - \frac{B \, \sigma}{20 \, m} \,.
\end{equation}
This expression indicates that the bending rigidity vanishes near the
critical point with the same exponent as the surface tension, i.e.
\begin{equation}
\label{eq:scaling}
k \propto \frac{B \, \sigma}{m} \, \propto \, \sigma \, d^2 \,.
\end{equation}
This scaling behavior should be contrasted to the usual
assumption that $k \!\propto\! \sigma \, \xi^2$, i.e. that
$k$ approaches a finite, non-zero limit at the critical point \cite{Meunier, Blokhuis93}. 

In Figure \ref{fig:k_and_sigma}, the gradient expansion result in
Eq.(\ref{eq:k_res}) for the bending rigidity is shown as the solid 
line. The bending rigidity is {\em negative}, in line with the simulation
results, although the magnitude is significantly lower.

To summarize, we have shown that to account for the simulated
scattering function over the whole range of scattering vector $q$, 
including the intermediate range between low $q$ (classical capillary
wave theory) and high $q$ (bulk-like fluctuations), one must 
take account of the interface's {\em bending rigidity}. Two of 
the important results are that the bending rigidity $k$ for 
the interface between phase-separated colloid-polymer 
mixtures is {\em negative}, and that on approach to the critical 
point it vanishes proportionally to the interfacial tension rather
than, as had often been supposed, varying proportionally to the
product of the tension and the square of the correlation length,
thereby approaching a finite, non-zero limit. Both features of $k$ 
are in accord with what is found in the simulations. The
magnitude of $k$ obtained from the molecular theory is lower
($\sqrt{-k/\sigma} \approx$~0.13~$d$) than in the
simulations ($\sqrt{-k/\sigma} \approx$~0.47~$d$; dashed line in
Figure \ref{fig:k_and_sigma}).


\begin{thebibliography}{99}

\bibitem{Daillant1}
C. Fradin, A. Braslau, D. Luzet, D. Smilgies, M. Alba, N. Boudet, K. Mecke, and J. Daillant,
{\it Nature} \textbf{403}.

\bibitem{Daillant2}
S. Mora, J. Daillant, K. Mecke, D. Luzet, A. Braslau, M. Alba,
and B. Struth, {\it Phys. Rev. Lett.} \textbf{90}, 216101 (2003).

\bibitem{Stecki}
J. Stecki and  S. Toxvaerd, {\it J. Chem. Phys.} \textbf{103}, 9763 (1995).

\bibitem{Vink}
R.L.C. Vink, J. Horbach, and K. Binder, {\it J. Chem. Phys.} \textbf{122}, 134905 (2005).

\bibitem{Tarazona}
P. Tarazona, R. Checa, and E. Chacon, {\it Phys. Rev. Lett.} \textbf{99}, 196101 (2007).

\bibitem{Evans}
R. Evans, {\it Adv. Phys.} \textbf{28}, 143 (1979).

\bibitem{RW} J.S. Rowlinson and B. Widom, {\it Molecular Theory of Capillarity}
(Clarendon, Oxford 1982).

\bibitem{BLS}
F.P. Buff, R.A. Lovett, and F.H. Stillinger, {\it Phys. Rev. Lett.} \textbf{15}, 621 (1965).

\bibitem{Weeks}
J.D. Weeks, {\it J. Chem. Phys.} \textbf{67}, 3106 (1977);
D. Bedeaux and J.D. Weeks, {\it J. Chem. Phys.} \textbf{82}, 972 (1985).

\bibitem{AOVrij}
S. Asakura and F. Oosawa, {\it J. Chem. Phys.} \textbf{22}, 1255 (1954);
A. Vrij, {\it Pure Appl. Chem.} \textbf{48}, 471 (1976).

\bibitem{Gast}
A.P. Gast, C.K. Hall, and W.B. Russel, {\it J. Coll. Interface Sci.} \textbf{96}, 251 (1983).

\bibitem{Lekkerkerker}
D.G.A.L. Aarts, M. Schmidt, and H.N.W. Lek\-ker\-ker\-ker, {\it Science} \textbf{304}, 847 (2004);
H.N.W. Lek\-ker\-ker\-ker, W.C.K. Poon, P.N. Pusey, A. Stroobants, and P.B. Warren,
{\it Europhys. Lett.} \textbf{20}, 559 (1992).

\bibitem{Parry}
A.O. Parry and  C.J. Boulter, {\it J. Phys. Condens. Matter} \textbf{6}, 7199 (1994).

\bibitem{Blokhuis99}
E.M. Blokhuis, J. Groenewold, and D. Bedeaux, {\it Mol. Phys.} \textbf{96}, 397 (1999).

\bibitem{Mecke}
K.R. Mecke and S. Dietrich, {\it Phys. Rev.} E. \textbf{59}, 6766 (1999).

\bibitem{Gibbs}
J.W. Gibbs, {\it Collected works} (Dover, NY, 1961).

\bibitem{Helfrich}
W. Helfrich, {\it Z. Naturforsch.} \textbf{28C}, 693 (1973).

\bibitem{Meunier}
J. Meunier, {\it J. Physique} \textbf{48}, 1819 (1987).

\bibitem{Varea}
C. Varea and A. Robledo, {\it Mol. Phys.} \textbf{84}, 477 (1995).

\bibitem{Kuipers}
M. Dijkstra, J.M. Brader, and R. Evans, {\it J. Phys. Cond. Matt.} \textbf{11}, 10079 (1999);
J. Kuipers and E.M. Blokhuis, {\it J. Coll. Interface Sci.} \textbf{315}, 270 (2007).

\bibitem{Blokhuis93}
E.M. Blokhuis and D. Bedeaux, {\it Mol. Phys.} \textbf{80}, 705 (1993).

\end{thebibliography}
\end{document}